\documentclass[twocolumn,prl,nofootinbib]{revtex4}
\usepackage{amsmath}
\usepackage{graphicx}
\usepackage{dcolumn}
\usepackage{bm}
\usepackage{amssymb}
\usepackage{latexsym}

\def\be{\begin{equation}}
\def\ee{\end{equation}}
\def\ba{\begin{eqnarray}}
\def\ea{\end{eqnarray}}

\begin{document}

\title{A Model Of Inflationary Cosmology Without Singularity}

\author{Yi-Fu Cai${}^{1}$}
\author{Taotao Qiu${}^{1}$}
\author{Jun-Qing Xia${}^{1,2}$}
\author{Xinmin Zhang${}^{1,2}$}

\affiliation{${}^{1}$Institute of High Energy Physics, Chinese
Academy of Sciences, P.O.Box 918-4, Beijing 100049, P.R.China}

\affiliation{${}^{2}$Theoretical Physics Center for Science
Facilities (TPCSF), Chinese Academy of Science, P.R.China}


\begin{abstract}

In this letter, we propose a model of inflationary cosmology with
a bounce preceded and study its primordial curvature
perturbations. Our model gives rise to a primordial power spectrum
with a feature of oscillation on large scales compared with the
nearly scale-invariant spectrum generated by the traditional slow
rolling inflation model. We will show this effect changes the
Cosmic Microwave Background (CMB) temperature power spectrum and
the Large Scale Structure (LSS) matter power spectrum. And further
with a detailed simulation we will point out this signal is
detectable to the forthcoming observations, such as PLANCK and
LAMOST.

\end{abstract}

\maketitle

Inflation, as a description of the very early universe, has
successfully resolved some problems existing in hot Big Bang
cosmology, such as flatness, horizon, monopole problem and so on
\cite{Guth:1980zm1}. However, this scenario is puzzled by the
initial singularity \cite{Borde:1993xh}. One possible approach to
this disaster is to introduce a bounce before the inflationary
expansion, which requires the hot Big Bang expansion be preceded
by a contracting period
\cite{Gasperini:1992em,Khoury:2001wf,Cai:2007qw}. If this happens,
one significant question would be proposed: what does bouncing
cosmology tell us for observations? Or, is it detectable for some
primordial relics from contracting phase to be imprinted on
observations? To answer this question, we need to study the
evolution of primordial gravitational perturbations seeded before
the bounce.

In this letter, we propose a nonsingular inflationary model, and
study signatures of its primordial fluctuations on CMB temperature
power spectrum and LSS matter power spectrum. We find an
interesting oscillation signature existing on large scales of the
scale-invariant spectrum and by a detailed simulation we will show
this new effect could be detected by the forthcoming astronomical
observations, such as PLANCK and LAMOST.

As in inflation theory, our model can be described in terms of
scalar fields which minimally couple to the four dimensional
Einstein's gravity. Explicitly it consists of two scalar fields
$\phi$ and $\psi$ with the lagrangian given by:
\begin{eqnarray}\label{lagrangian}
{\cal L}=
\frac{1}{2}\partial_{\mu}\phi\partial^{\mu}\phi-\frac{1}{2}\partial_{\mu}\psi\partial^{\mu}\psi
-V(\phi,\psi)~,
\end{eqnarray}
in a spatially flat Friedmann-Robertson-Walker (FRW) universe.
Here the essential component is the scalar field $\psi$. It plays
a crucial role in giving a bouncing solution smoothly. Without it,
the model in Eq. (1) will be similar to the traditional inflation
with a single scalar field, which as we know suffers from the
problem of the initial singularity. In this model the potential is
only the function of the field $\phi$ and of Coleman-Weinberg form
\cite{Coleman:1973jx}:
\begin{eqnarray}
V=\frac{1}{4}\lambda\phi^4 \left( \ln\frac{|\phi|}{v}-\frac{1}{4}
\right) + \frac{1}{16}\lambda v^4~,
\end{eqnarray}
which takes its maximum value $\lambda v^4/16$ at $\phi=0$ and
vanishes at the minima when $\phi=\pm v$. Therefore, the scalar
field $\psi$ merely affects the evolution around the bounce but
decays out quickly when away from it.

In order to discuss the perturbations explicitly, we first see how
the background universe evolves. In this model a contracting
universe can be driven to reach a minimal size during which the
universe evolves like a matter-dominant one, and then a
quasi-exponential expansion is following, and so is able to
explain the problems appeared in standard Big Bang cosmology. The
process to link the contraction and expansion is a smooth bounce,
and the evolution of the hubble parameter can be treated as a
linear function of the cosmic time approximately.

We take the initial condition for the background as that $\phi$
stays at one vacuum like $-v$ when the universe is contracting and
$\dot\psi$ is small enough which can be ignored on background
evolution. In this phase, the field $\phi$ oscillates around $-v$
making the equation-of-state (EoS) of the universe oscillate about
$w=0$, and so the average state being similar to a
matter-dominated one. Thus we have the useful expressions of
background evolution
\begin{eqnarray}\label{relationc1}
a\sim(-\eta)^{2}~,~~{\cal
H}=\frac{2}{\eta}~,~~|\dot\phi|\sim\eta^{-3}~,
\end{eqnarray}
where ${\cal H}\equiv{a'}/{a}$ is the comoving hubble parameter
and the prime denotes the derivative with respect to the comoving
time. Another useful relation is given by
\begin{eqnarray}\label{relationc2}
\frac{\phi''}{\phi'}=\frac{2{\cal H}{\cal H}'-{\cal H}''}{2({\cal
H}^2-{\cal H}')}~,
\end{eqnarray}
which will be used to calculate the metric perturbations.

Since the universe is contracting, the amplitude of the field
$\phi$'s oscillation gets larger and larger, while the
contribution of the field $\psi$ grows rapidly. When the field
reaches the plateau, the bounce happens at the moment $t_{B-}$.
During the bounce, we take the parametrization
$H(t)=\alpha(t-t_B)$ around the bounce point $t_B$, and the
coefficient $\alpha$ is a positive constant determined from
numerical calculations. In the bouncing phase, the kinetic term of
$\psi$ reaches the maximal value and from the equation of motion
we deduce an expression
$\ddot\phi/\dot\phi=-3H\dot\psi^2/(\dot\psi^2-\frac{\alpha}{4\pi
G})\simeq -3H$ when $\alpha$ is not very large. Finally we have
the approximate relations
\begin{eqnarray}\label{relationb}
{\cal H}\simeq\frac{y}{2}(\eta-\eta_B),~\phi''\simeq-2{\cal
H}\phi',~|\dot\phi|\sim e^{-\frac{3}{4}y(\eta-\eta_B)^2}~,
\end{eqnarray}
where we have defined $y\equiv8\alpha a_B^2/\pi$.

\begin{figure}[htbp]
\includegraphics[scale=0.4]{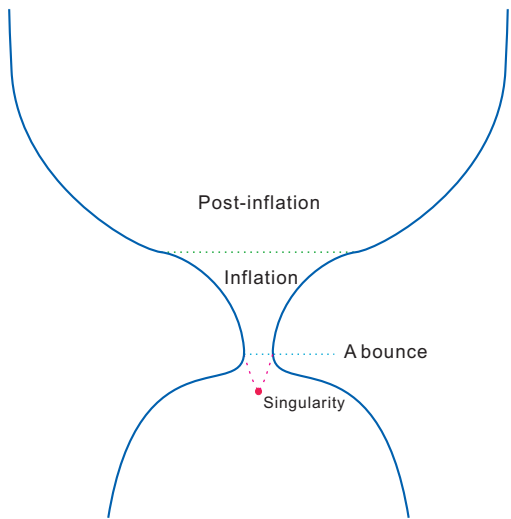}
\caption{\textbf{Evolution of the universe.} A sketch plot of the
evolution of the universe in the model of Eq.(\ref{lagrangian}).
Before inflation, there is a bounce instead of the initial
singularity.} \label{fig:scalefactor}
\end{figure}

\begin{figure}[htbp]
\includegraphics[scale=0.7]{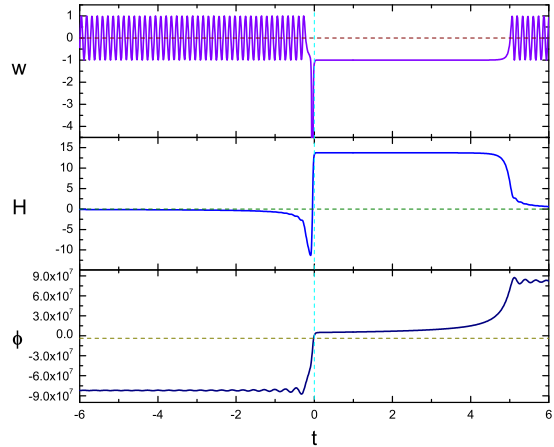}
\caption{\textbf{Evolutions of $\phi$, $w$ and $H$.} A plot of the
evolutions of the field $\phi$, the EoS $w$ and the hubble
parameter $H$ in the model of Eq.(\ref{lagrangian}) where we take
the Coleman-Weinberg potential. In the numerical calculation we
choose the parameter $\lambda=8.0\times10^{-14},~v=0.82M_{pl}$,
and the initial condition as:
$\phi=-0.82M_{pl},~\dot\phi=3.0\times10^{-10}M_{pl}^2,~\psi=-0.72M_{pl},~\dot\psi=5.0\times10^{-13}M_{pl}^2$
where $M_{pl}\equiv1/\sqrt{G}$~.} \label{fig:background}
\end{figure}

After the bounce, as the field $\phi$ moves forward slowly along
the plateau, the universe enters into an expanding phase at the
moment $t_{B+}$ and the EoS of the universe is approximately $-1$.
The universe expands with its scale factor growing almost
exponentially. In this phase, we have the well-known relations for
background evolution
\begin{eqnarray}\label{relationi}
a\sim-\frac{1}{\eta}~,~~H\sim \mathrm{Constant}~.
\end{eqnarray}
Finally, when the field drops into the vacuum $+v$, it will
oscillate again and the EoS of the universe will oscillate around
zero as it does before the bounce. To make the scenario explicit,
we give a sketch description of our model in
Fig.\ref{fig:scalefactor} and present the numerical calculation of
the background parameters in Fig.\ref{fig:background}.

Now we study the linear perturbations of the model. Taking the
longitudinal (conformal Newtonian) gauge, the metric perturbation
is presented as follows:
\begin{eqnarray}
ds^2=a^2(\eta) \left[(1+2\Phi)d\eta^2-(1-2\Psi)dx^idx^i \right]~,
\end{eqnarray}
and the equation of motion of the gravitational potential is:
\begin{eqnarray}\label{eomp}
\Phi''+2({\cal H}-\frac{\phi''}{\phi'})\Phi'+2({\cal H}'-{\cal
H}\frac{\phi''}{\phi'})\Phi-\nabla^2\Phi\nonumber\\=8\pi G (2{\cal
H}+\frac{\phi''}{\phi'})\psi'\delta\psi~,
\end{eqnarray}
which can be derived from the basic perturbation equations
directly (we refer the complete derivation to
Ref.\cite{Cai:2007zv}, and see e.g. Ref.\cite{Mukhanov:1990me} for
a comprehensive survey of the cosmological perturbation theory).
As is pointed out previously, the energy density of the field
$\psi$ is usually negligible far away from the bounce, and hence
we have $\psi'\simeq 0$. Near the bounce, $\psi$ becomes very
important, but according to the analysis in Ref.\cite{Cai:2007zv}
we have the approximation $2{\cal H}+\phi''/\phi'\simeq 0$ and so
the perturbation of $\psi$ decouples from Eq. (\ref{eomp}).
Therefore, we will neglect the r.h.s. of Eq. (\ref{eomp}), and
just focus on the adiabatic fluctuations in the following which
can be determined by a single scalar field $\phi$.

\begin{figure}[htbp]
\includegraphics[scale=0.3]{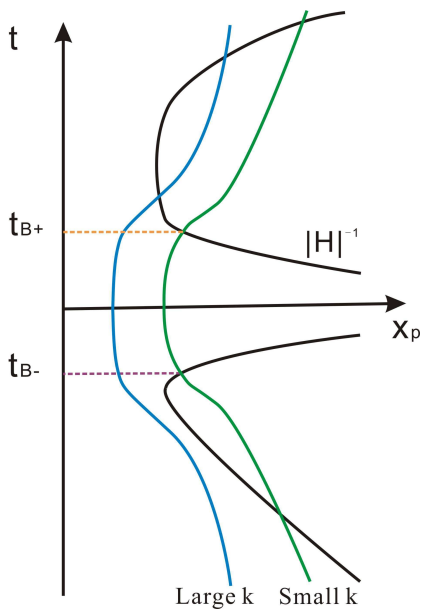}
\caption{\textbf{Evolution of perturbations.} A sketch plot of the
evolution of perturbations with different comoving wave number $k$
in our model.} \label{fig4:sketch}
\end{figure}

Now we follow one Fourier mode of the perturbation, labelled by
its comoving wave number $k$, and find that there are two paths
for perturbations. The evolution of perturbations is sketched in
Fig.\ref{fig4:sketch}. Initially all the perturbations stay inside
the horizon in the far past. Since the hubble radius shrinks,
those modes with small comoving wave number exit the horizon while
the large $k$ scales still keep inside. When the bounce takes
place, all the perturbations will enter the horizon because at
that moment the hubble radius diverges. Since in our model the
bounce is followed by an expanding phase, those Fourier modes will
escape out if the efolds for the post-bounce slow-rolling period
is large enough. After that, these modes will re-enter the horizon
at late times after the slow-rolling phase has finished. In the
following calculations, we will focus on large $k$ region and see
whether the large k modes are able to perform scale-invariant
spectra and give more information on the CMB observations.

For the contracting phase before the bounce, the equation of
motion in momentum space can be solved explicitly. We take the
Bunch-Davies vacuum as the initial condition $\Phi_k\sim\frac{4\pi
G}{\sqrt{2k^3}}|\dot\phi|e^{-ik\eta}$ when the perturbation are
deeply inside the horizon. Since during this period
$|\dot\phi|\sim\eta^{-3}$, we obtain
\begin{eqnarray}\label{sol1}
\Phi_k=4\pi
G\frac{\sqrt{\rho_i}\eta_i^3}{\eta^3}\frac{e^{-ik\eta}}{\sqrt{2k^3}}~,
\end{eqnarray}
where the subscript $i$ represents the initial time. Substituting
the Eq. (\ref{relationb}) into Eq. (\ref{eomp}) and solving it, we
have the solution to the perturbation in the bouncing phase
\begin{eqnarray}\label{sol2}
\Phi_k &\simeq& e^{-\frac{3}{4}y(\eta-\eta_B)^2} \nonumber\\
 &\times& \bigg\{ C_k\cos[k(\eta-\eta_B)] +D_k\sin[k(\eta-\eta_B)]
\bigg\}.
\end{eqnarray}
Moreover, for the nearly de-Sitter expanding phase, we obtain
\begin{eqnarray}\label{sol3}
\Phi_k &=& (\eta-\tilde\eta_{B+})^{\gamma} ~ [ k^{-\nu} E_k
J_{\nu}(k(\eta-\tilde\eta_{B+})) \nonumber\\
&+& k^{\nu} F_k J_{-\nu}(k(\eta-\tilde\eta_{B+})) ]~,
\end{eqnarray}
where $\gamma\simeq1/2$, $\nu\simeq1/2$ and
$\tilde\eta_{B+}\equiv\eta_{B+}+1/{\cal H}_{B+}$.

Having obtained the solutions of the perturbation equations in
different phases, it is necessary to know the matching relations
among these solutions and determine the coefficients $C_k$, $D_k$,
$E_k$ and $F_k$. This depends on whether the curvature
perturbation on a uniform comoving hypersurface or the
gravitational potential passes through the bounce regularly
\cite{Hwang:1991an} (see also \cite{Deruelle:1995kd} for a recent
study). For a nonsingular bounce scenario such as what we
considered, the continuity of background evolution implies that
both $\Phi$ and $\Phi'$ are able to pass through the bounce
smoothly. By matching $\Phi$ in Eqs. (\ref{sol1}) and (\ref{sol2})
on the surface $\eta_{B-}$, and that in Eqs. (\ref{sol2}) and
(\ref{sol3}) in sub-hubble region on the surface $\eta_{B+}$ as
well as their comoving time derivatives, all of the coefficients
can be determined. However, since $E_k$ represents a decaying mode
when escape outside the horizon, we neglect it and finally obtain
the dominant mode
\begin{eqnarray}\label{Fk}
F_k
 &\simeq& \sqrt{\frac{\pi}{2}} \frac{4\pi
 G}{\sqrt{2k^3}}|\dot\phi| e^{-ik\tilde\eta_{B+}} \nonumber\\
&\times&
\left\{1+\frac{3e^{-ik(\eta_{B-}-\tilde\eta_{B+})}}{k\eta_{B-}}
\sin[k(\eta_{B-}-\tilde\eta_{B+})] \right\}.
\end{eqnarray}

By comparing the coefficient of Eq. (\ref{Fk}) and the initial
form of $\Phi$ in Eq. (\ref{sol1}), one obviously notice that the
sub-hubble form of the metric perturbation has obtained an
oscillation term when the universe undergoes a bounce. Note that,
another important quantity is the curvature perturbation in
comoving coordinate $\zeta\equiv\Phi+\frac{{\cal H}}{{\cal
H}^2-{\cal H}'}( \Phi'+{\cal H}\Phi )$, and when the expansion is
nearly de-sitter like, there is a simple relation between these
two quantity $\zeta \simeq \Phi/\epsilon$ with the slow roll
parameter $\epsilon\equiv-\dot H/H^2$. Therefore, we eventually
have the primordial power spectrum for the curvature perturbation
\begin{eqnarray}\label{Pzeta}
P_{\zeta}\simeq\frac{8}{3}G^2\frac{\rho}{\epsilon}\bigg\{
1-\frac{3{\cal H}_{B-}}{2k}\sin\frac{2k}{{\cal H}_{B+}} \bigg\}~,
\end{eqnarray}
where we have assumed $\eta_{B+}-\eta_{B-}$ to be small compared
with $|\frac{1}{{\cal H}_{B+}}|$. Obviously, the first term
provides a nearly scale-invariant spectrum which is consistent
with current cosmological observations. However, the second term
apparently shows that there is a wiggle on the spectrum, due to
the modification brought by a bounce.

\begin{figure}[htbp]
\includegraphics[scale=0.4]{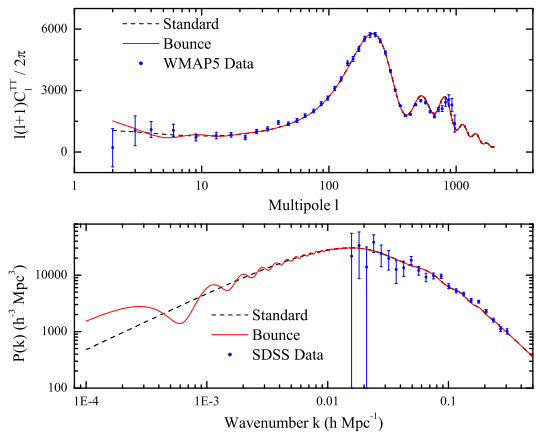}
\caption{\textbf{Observational effects by a bounce.} The effect on
the CMB temperature power spectrum and LSS matter power spectrum
by our model ${\cal H}_{B+}\simeq2\times10^{-4}\mathrm{Mpc}^{-1}$
and ${\cal H}_{B-}\simeq-1.6\times10^{-4}\mathrm{Mpc}^{-1}$. The
dots and error bars are WMAP5 and SDSS data.} \label{fig3:Bounce}
\end{figure}

For a numerical estimate of ${\cal H}_{B+}$ and ${\cal H}_{B-}$,
we normalize the current scale factor $a_0=1$ and choose the
current hubble parameter
$H_0=72~\mathrm{km~s^{-1}~Mpc^{-1}}=1.536\times10^{-42}\mathrm{GeV}$,
the hubble parameter during inflation $H_i\simeq 1.68\times
10^{12}\mathrm{GeV}$, and the e-folds for inflation $N\simeq60$.
Therefore, our model predicts that ${\cal
H}_{B+}\simeq2\times10^{-4}\mathrm{Mpc}^{-1}$ and ${\cal
H}_{B-}\simeq-1.6\times10^{-4}\mathrm{Mpc}^{-1}$. Based on the
primordial spectrum in Eq. (\ref{Pzeta}), in Fig.\ref{fig3:Bounce}
we illustrate the CMB temperature power spectrum and present LSS
matter power spectrum. For comparison we have also considered the
standard case where ${\cal H}_{B-}$ is taken to be zero. In the
numerical calculations we use the publicly available Markov Chain
Monte Carlo (MCMC) package CosmoMC \cite{cosmomc} and take the
basic cosmological parameters as given below:
\begin{eqnarray}
&&(\Omega_b h^2,\Omega_c h^2, \tau, H_0, A_s)\nonumber\\
&=&(0.022,0.115,0.088,72,2.3\times10^{-9})~,
\end{eqnarray}
and the pivot scale $k_{\ast}=0.05~\mathrm{Mpc^{-1}}$. One can see
from Fig.\ref{fig3:Bounce} that our model leads to an obvious
$k$-dependent oscillation signature in the power spectrum,
especially at large scales.

To test our model we firstly consider the current astronomical
observations from  WMAP5 \cite{Komatsu:2008hk} and SDSS
\cite{Tegmark1}, due to the large uncertainties at large scales as
shown in Fig.\ref{fig3:Bounce} by the blue dots and the error
bars, we find that the oscillating spectrum of our model is
consistent with the data. Thus we consider the forthcoming
measurements PLANCK \cite{planck} and LAMOST \cite{lamost} with
higher precision. We simulate the CMB TT, TE and EE power spectra
with the sensitivity of PLANCK and the LSS linear matter power
spectrum with the sensitivity of LAMOST and find these
measurements will be sensitive to ${\cal
H}_{B-}\simeq-7.0\times10^{-5}\mathrm{Mpc}^{-1}$ which is smaller
than our predicted value and makes our model testable. If this
signal would be detected, it will act as a ${\it smoking~gun}$ to
the bouncing cosmology.

Physics of bouncing cosmology, since it happens in extremely high
energy regime, is hardly to be found by experiments directly. So
it is a debate whether a bounce has taken place or not. To find
the evidences of a bounce, we need to know what can a bounce leave
for observations. This question is still discussed drastically in
the literature, and one potential clue is to study the primordial
curvature fluctuations. In the context of the Pre-Big-Bang
scenario \cite{Gasperini:1992em} and in the cyclic/Ekpyrotic
cosmology \cite{Khoury:2001wf}, the resulting curvature
perturbation strongly depends on the physics at the epoch of
thermalization, and thus an uncertainty of a thermalized surface
is involved \cite{Brustein:1994kn,Lyth:2001pf1}. In the frame of
loop quantum cosmology, it is argued that fluctuations before and
after the bounce are largely independent \cite{Bojowald:2007zza}
(yet see Ref.\cite{Corichi:2007am} for some criticisms). In this
letter we propose a concrete cosmological model with inflation
preceded by a bounce, and by investigating in detail the
perturbations we show some imprints of the bounce are detectable
to the forthcoming CMB and LSS observations.

\section*{Acknowledgments}

We have performed our numerical analysis in the Shanghai
Supercomputer Center (SSC). We thank Robert Brandenberger, Mingzhe
Li and Paul Steinhardt for useful comments on the manuscript. This
work is supported in part by National Science Foundation of China
under Grant No. 10533010, and the 973 program No.2007CB815401, and
by the Key Grant Project of Chinese Ministry of Education (No.
305001).

\end{document}